\documentclass[letter]{aa} 

\usepackage{graphicx}
\usepackage{txfonts}
\usepackage{color}
\newcommand{\om}[1]{{\color{black} #1}}

\begin{document}

   \title{Hunting ghosts: the iconic stellar stream(s) around NG5907 under scrutiny}

   \author{Oliver M\"uller
          \inst{1}
          \and
          Ana Vudragovi\' c\inst{2}
          \and
          Michal B\'ilek\inst{1}}

   \institute{Observatoire Astronomique de Strasbourg  (ObAS),
Universite de Strasbourg - CNRS, UMR 7550 Strasbourg, France\\
 \email{oliver.muller@astro.unistra.fr}
         \and
             Astronomical Observatory, Volgina 7, 11~060 Belgrade, Serbia\\
             }

   \date{Received tba; accepted tba}

 
  \abstract
   {Stellar streams are  regarded as crucial objects to test galaxy formation models, with their morphology tracing the underlying potentials and their occurrence tracking the assembly history of the galaxies. The existence of one of the most iconic stellar streams, the double loop around NGC\,5907, has recently been questioned by new observations with the Dragonfly telescope. This new work only finds
   parts of the stream, even though they reach a $1\sigma$ surface brightness limit of 30.3\,mag arcsec$^{-2}$ in the $g$-band. 
   Using 7.2 hours of \om{Luminance} $L$-band imaging with the Milankovi\'c 1.4 meter telescope, we have re-observed the putative double loop part to confirm or reject this assessment. We do not find signs of the double loop, but see only a single, knee-shaped stellar stream.
   Comparing our to the data by the Dragonfly team, we find the same features. Our observations reach a $1\sigma$ surface brightness limit of 29.{7}\,mag arcsec$^{-2}$ in the $g$-band. \om{These findings emphasize the need for independent confirmation of detections of very low-surface brightness features.} 
   }
  
               
   \maketitle
%

\section{Introduction}
Over the last years, deep imaging of nearby galaxies have revealed a plethora of low-surface brightness features, such as dwarf galaxies, stellar streams, shells, and tidal debris. Such features unlock crucial information for galaxy formation and are regarded as tests for the current cosmological model ($\Lambda$CDM), in which galaxies are formed in a hierarchical fashion by accretion of smaller galaxies (\citealt{2005ApJ...635..931B}{, but see also \citealt{2018A&A...614A..59B,2019arXiv190807537B}}).
Some fossils of these ancient accretion events are expected to be detectable today, with the number of these ghosts and their phase-space correlation being a strong test for the $\Lambda$CDM model of galaxy formation \citep[e.g., ][]{2002MNRAS.332..915I,2017MNRAS.470...60E,2018ApJ...867..101B,2019ApJ...883...87P}. 

In the Local Group, it is known for a long time that streams wind around the Milky Way \citep[e.g., ][]{1997AJ....113..634I,2007ApJ...658..337B} and the Andromeda galaxy \citep{2014ApJ...787...19M,2018ApJ...868...55M}, based on star count studies. With the Gaia mission, the number of streams detected around the Milky Way has exploded \citep{2018MNRAS.481.3442M,2019ApJ...872..152I}. Outside of our Local Group, the detection of resolved stellar streams is an almost impossible task with current instrumentation, with the stream around NGC\,891 \citep{2010ApJ...714L..12M} at 10\,Mpc setting the record of what is achievable today. At 
{larger}
distances, only integrated light observations are able to uncover these fossils. One of the most iconic stream discovered by integrated light observations is found around NGC\,5907 at 14\,Mpc \citep{1998ApJ...504L..23S}, where \citet{2008ApJ...689..184M} uncovered a double looping structure, which was not visible in the  \citet{1998ApJ...504L..23S} data. Noteworthy, follow-ups of this stream with Subaru/Suprime Cam \citep{2016AJ....152...72L} detected only the parts which were already visible in \citet{1998ApJ...504L..23S}, but not the double loop.  The observation by \citet{2008ApJ...689..184M} was done with a modest telescope of only 0.5 meter diameter in size and was conducted by dedicated amateur astronomers. 
Since then, many more stellar streams around other giant galaxies have been detected by this amateur astronomer team, see e.g. \citet{2010AJ....140..962M}.

The search for stellar streams is not only conducted by amateurs. Several professional teams have taken up the efforts to find streams and shells around nearby galaxies \citep[e.g., ][]{2014ApJ...791...38W,2015ApJ...800L...3W,2015MNRAS.446..120D,2016A&A...588A..77B,2016ApJ...823...19C,2019ApJ...883L..32V}. One of these teams, using the Dragonfly telescope \citep{2019arXiv191014045D}, has re-observed the iconic structure around NGC\,5907 and to their surprise, found only part of the stream(s) {reported by \citet{2008ApJ...689..184M}}, with no trace of the double loop \citep{2019ApJ...883L..32V}. Furthermore, they found a missmatch in the positions of the detected parts of the stream  compared to those provided  by \citet{2008ApJ...689..184M}.  
{This calls for an independent observation.}
To address this issue, we observed the putative double loop part of NGC\,5907 with the newly commissioned Milankovi\'c 1.4 meter telescope \citep{2018POBeo..98..333S} located at the top of the Vidojevica mountain, Serbia. 
{The results are presented in this letter.}
\begin{figure*}
    \centering
    \includegraphics[width=18cm]{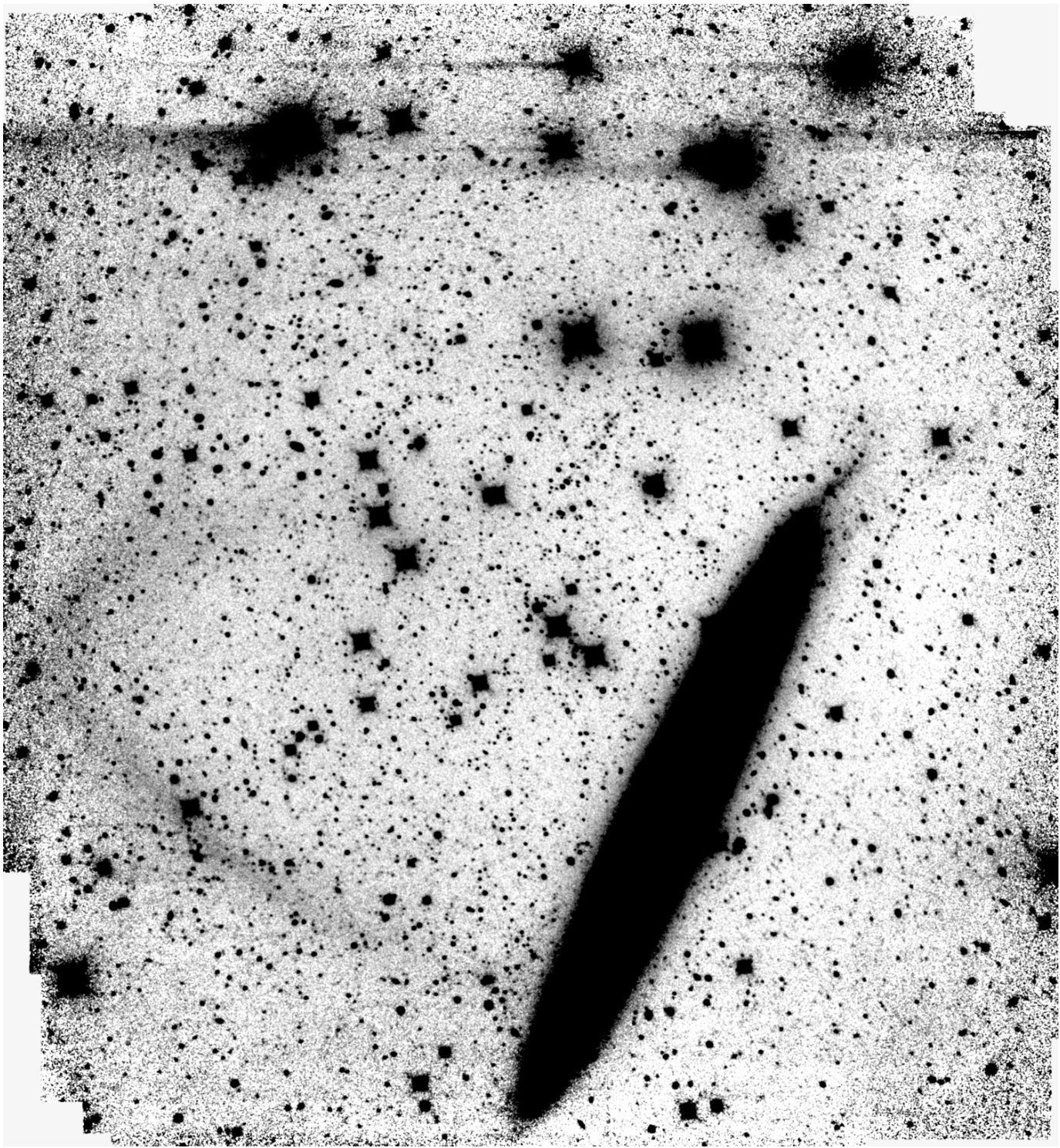}
    \caption{The fully calibrated and stacked image of NGC\,5907. The image was slightly enhanced with a Gaussian blur. The image is $18\farcm4\times20\farcm2$ in size. North is to the top, east to the left.}
    \label{fig:image}
\end{figure*}

\section{Observations and data reduction}
The galaxy NGC\,5907 was observed 
on three consecutive nights  (24 - 26 October 2019)
with the 1.4\,meter Milankovi\' c telescope mounted at the Astronomical Station Vidojevica (Serbia), using an Andor IKONL CCD camera. {The observations were taken at an airmass between $1.66 < X < 2.97$.  The dark current is 0.02 e-/sec.}
All images were taken in the luminance $L$-filter with {300 second exposures.} 
The integrated exposure time reached was 7.2\,hrs. We had photometric nights with an average seeing of 1\farcs4. The galaxy was at an altitude of 40 degrees at the beginning and went down to 20 degrees at the end of observations each night. Utilizing the focal reducer delivers a square field-of-view with sides of 13\farcm3. {We applied a large dithering pattern randomized within a maximum offset of 4\arcmin, which is crucial when digging deep into the low-surface brightness regime at the 28-29\,mag arcsec$^{-2}$ level \citep{2009PASP..121.1267S,2019arXiv190909456M} and beyond \citep{2016ApJ...823..123T}. Such a strategy provides a strong handle on the background subtraction, as well as a removal of systematics arising from the instruments and observations, i.e. from reflections or flat-fielding artefacts.} 

The calibration of the images follows the standard procedure. 
{A master bias frame was created from ten single bias frames, using a median stack. A master dark frame was created from ten single dark frames, using again a median stack. These master dark and bias frames were} then subtracted 
{from} all individual flat fields and scientific exposures. A master flat field was made from 30 single {twilight} flat fields, which have been normalized by their median value. The stack was again made with a median. The resulting master flat field {was flat to the 1-2 percent level and} was then applied to all individual dark-subtracted scientific exposures.

For the surface brightness limits we aimed to reach, it is crucial to carefully model the background. Approaches of background modelling, like it is done in Source Extractor \citep{1996A&AS..117..393B}, are prone to removing low-surface brightness features as part of the sky. Therefore, we took a custom approach. On every previous calibrated scientific exposure, we ran a source detection using Source Extractor with a 1$\sigma$ threshold, {creating a segmentation map, which was then applied as} a mask on the scientific exposures. {Because Source Extractor will miss the faint outskirts of astronomical objects, we additionally dilated the mask with a 5$\sigma$ gaussian kernel. The resulting masked image} then only contains pixels corresponding to the background. The masked scientific exposures were then normalized by their median values, and stacked together. This stack will give a background model, which is free from astronomical sources due to the large dithering pattern we have applied. Because it is normalized, it can be scaled to the measured background values in the scientific exposures using the {the global background value estimated by Source Extractor} and ultimately removed, resulting in our final fully calibrated and background subtracted science frames. The final, fully calibrated and dithering image reached a field-of-view of 18\farcm8 $\times$ 20\farcm3. {The python pipeline can be found on gitlab\footnote{{gitlab.com/VoltarCH/milankovic-telescope-reduction-pipeline}}.}

The astrometry in the science frames was solved by astrometry.net \citep{2010AJ....139.1782L}. To create our final scientific image, we have stacked all science frames with IRAF's \textit{imcombine} command, using a median stack and an average sigma clipping rejection algorithm.

The zero point of this stacked image was estimated using the Pan-STARRS DR2 star catalog \citep{2016arXiv161205242M}. We used DAOPHOT2 \citep{1987PASP...99..191S} aperture photometry to measure the instrumental 
$L$-band magnitudes of the stars.
There are no $L$-band star catalogs yet available, therefore, we needed to transform the $L$-band to the more commonly used SDSS bands. \citet{2016A&A...588A..89J} successfully performed a linear transformation between the $L$-band and the $r$-band and found only a minor dependence on the $g-r$ color (see their Figure 1). Therefore, we have calibrated our image to $g$-band standard stars {using $g=c_0L+c_1$, based on 78 standard stars in a magnitude range between 17 and 22. The transformation is given with $c_0=1.03$ and $c_1=32.12$\,mag.} 
The magnitude system of Pan-STARRS is in AB.

{For the parts of the image with 100\% coverage, the surface brightness limit was measured from the standard deviation in $10\times10$ arcsec$^2$ boxes of empty pixels, meaning that they don't contain any (visible) stars.}
We estimate a $1\sigma$ surface brightness limit of $29.7\pm0.04$ mag arcsec$^{-2}$ in the $g$-band (corresponding to a $3\sigma$ limit of $28.4\pm0.04$ mag arcsec$^{-2}$). \om{The error has been estimated by the standard deviation of the measured surface brightness of 14 such boxes.}

\section{Low surface brightness features}
In Figure\,\ref{fig:image} we present the fully stacked and calibrated image of the surroundings of NGC\,5907, slightly enhanced with a Gaussian blur. What low-surface brightness features does it reveal?
We detect a narrow low-surface brightness feature stretching 10\farcm5 ($\approx43$\,kpc at 14\,Mpc) from the disk of the galaxy towards the east and then bending 4\farcm2 ($\approx17$\,kpc at 14\,Mpc) towards the north, where it gradually disappears. This morphology is reminiscent of stellar streams found in other galaxy group, e.g. around NGC\,1052 \citep{2019A&A...624L...6M}.

\begin{figure}
    \centering
    \includegraphics[width=8.5cm]{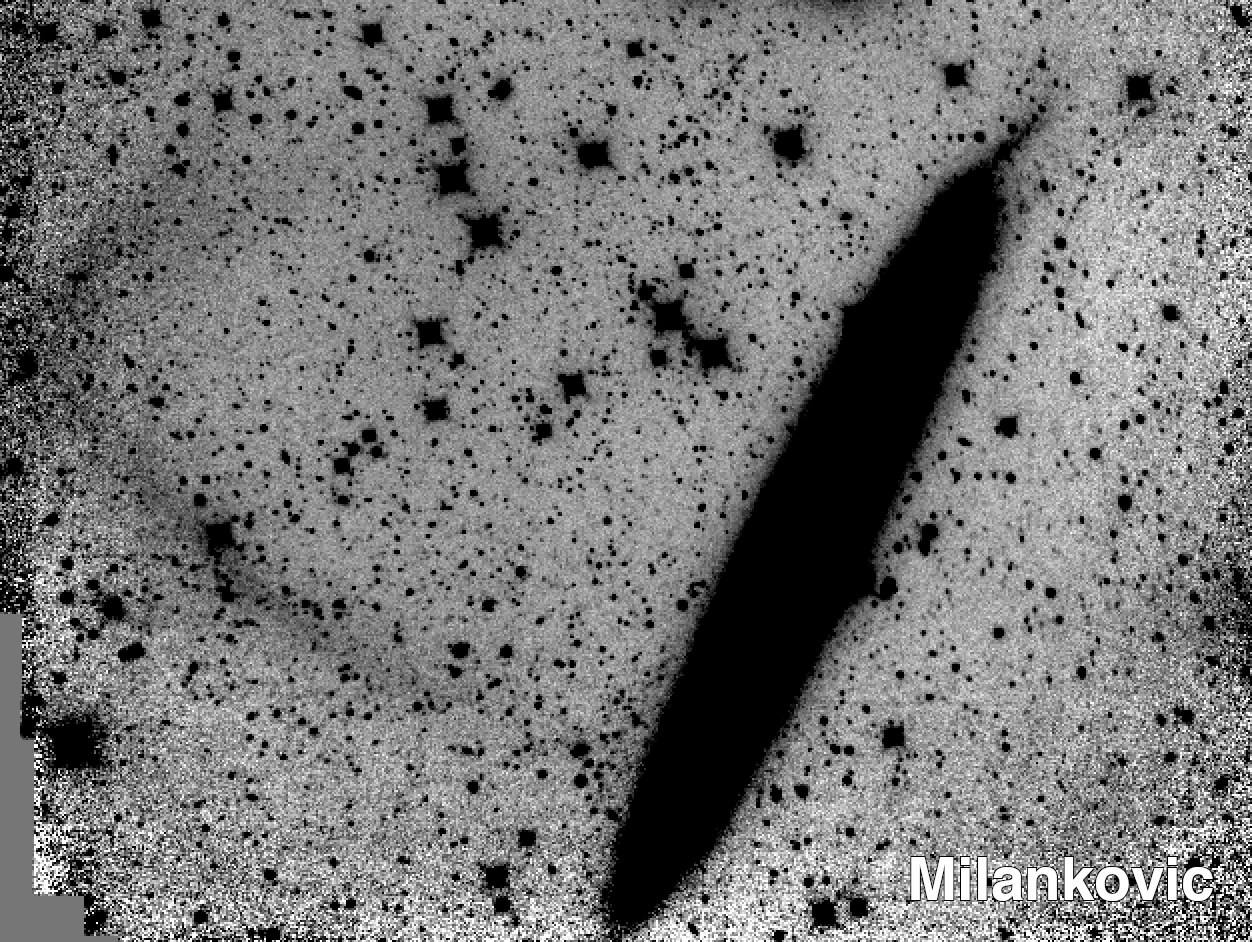}\\
    \includegraphics[width=8.5cm]{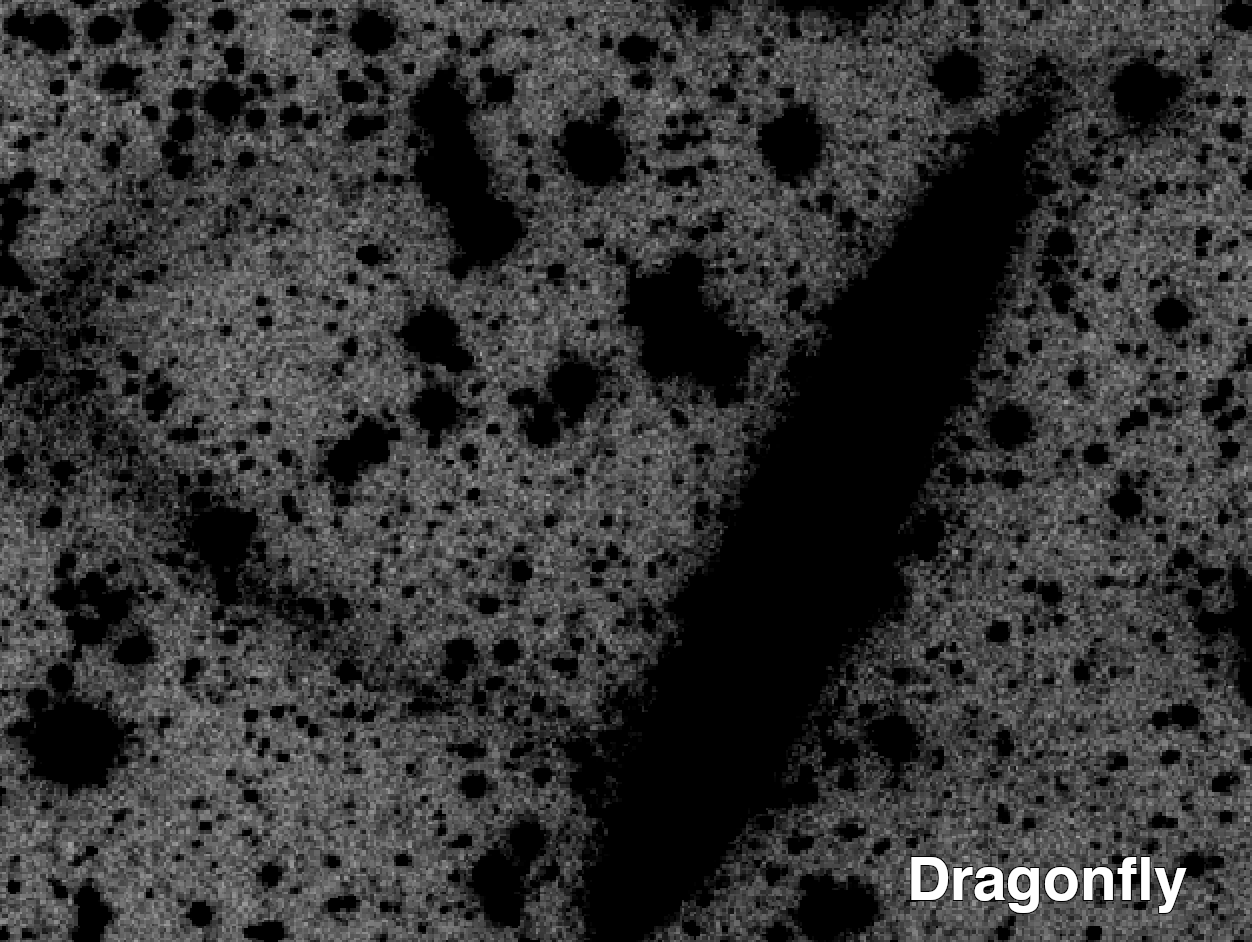}
    \caption{Comparison between our observations (top) and the relevant part of the Dragonfly image (bottom), kindly provided by the Dragonfly team.}
    \label{fig:comparison}
\end{figure}

\begin{figure*}
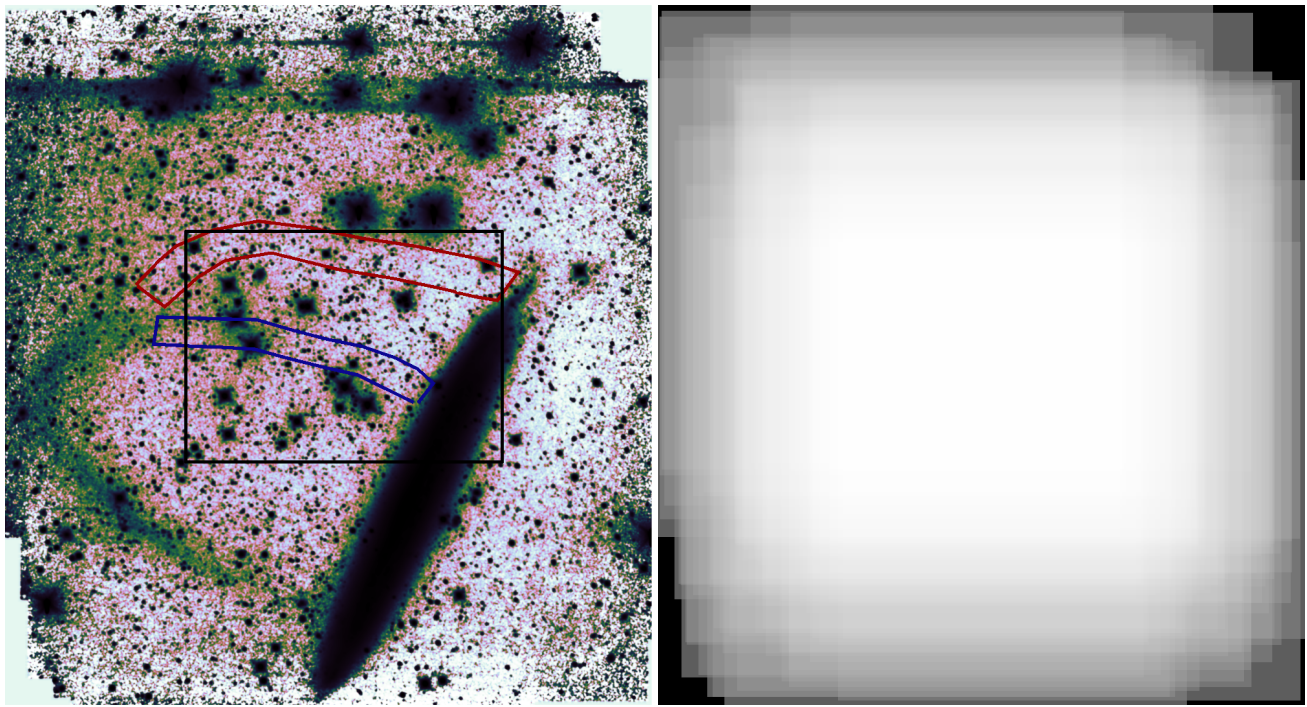

    \centering
    \includegraphics[width=8.5cm]{galaxy_enhanced2.png}
    \includegraphics[width=8.5cm]{weightmap_log.png}
    \caption{Left: The image enhanced with a Gaussian blur and a histogram equalization as discussed in \citet{2008ApJ...689..184M}. The paths indicate the positions of the missing parts of the streams east of the galaxy, this is, the iconic part of the double loop system. {The black box corresponds to the region where a 100\% coverage has been achieved. Right: Weight map of the stack {in a logarithmic representation}. White corresponds to a 100\% coverage, i.e. the area with 7.2 hour exposure in total, black to 0\%.}}
    \label{fig:parts}
\end{figure*}

In Figure\,\ref{fig:comparison} we compare the relevant part of our image with the Dragonfly $g$-band image taken by \citet{2019ApJ...883L..32V}. The shape of the stream detected by us and by \citet{2019ApJ...883L..32V} 
are identical, which is also apparent when 
blinking the two images. So where should the double loop reside? In Figure\,\ref{fig:parts} we indicate the 
{contours of the double stream reported by \citet{2008ApJ...689..184M}. There are no hints of these structures in our image.
We applied the same post-processing techniques as} \citet{2008ApJ...689..184M}, namely, a histogram equalization and a Gaussian blur. 
Along these regions, we measure a $1\sigma$ surface brightness limit of 29.7\,mag arcsec$^{-2}$ $g$-band (i.e. a $3\sigma$ limit of 28.4\,mag arcsec$^{-2}$). However, we note that this is measured before the image is enhanced{, giving a lower limit of what is detectable.}
{In Figure\,\ref{fig:parts} we also show the part of the image with a 100\% coverage of exposures, corresponding to the deepest part of the full stack. The two putative streams are residing well within this area, so incompleteness of the data should play no role. In contrast, the part of the stream to the west (i.e. bottom right) of NGC\,5907 is not detectable in our image, whereas it is visible in the Dragonfly images (see Figure\,\ref{fig:comparison}). This part coincides with an incomplete area of the image, see also our weight map in Figure\,\ref{fig:parts}.}

{The faintest stars detected in our images are of the order of $25.8$ to $26.0$\,mag in the $g$-band, which is several magnitudes deeper than the detection limit of e.g. SDSS. The magnitudes were estimated with simple aperture photometry on several hand-picked faint stars.}



\section{Discussion and conclusion}
The recent paper by \citet{2019ApJ...883L..32V} has stirred a controversy which has been hotly debated at conferences and on social media. To shed light on the discussion, we have independently taken deep data using a 1.4 meter telescope to assess whether there is a double loop structure as found by \citet{2008ApJ...689..184M} or rather a simple knee-shaped stream as found by \citet{2019ApJ...883L..32V}. Our imaging is consistent with the latter, with no signs of a double loop. 

We cannot assess whether there was a problem with the data reduction by \citet{2008ApJ...689..184M}, because the data is unfortunately not available. Therefore, we can only rely on re-observations of the system.
An argument raised in favor of the existence of a double loop is the fact that many amateur astronomers have repeatedly detected it. 
\om{However, the data reduction procedures adopted for the amateur images are not always transparent. The fact that now two completely independent professional teams could not confirm the double loop puts some doubts on at least some of the previously found very low-surface brightness features. The limiting surface brightness of amateur images is rather difficult to assess.}
Both our data and the data by \citet{2019ApJ...883L..32V} have applied large dithering patterns,
which indicates that {the handling of the data could be one of the culprits \om{ causing the disagreement}.} In this respect it is also striking that in the deep Subaru/Suprime Cam data by \citet{2016AJ....152...72L} no double loop feature was visible, even though it was not discussed in their work. {It seems that professionally handled data always yields a single stream, while data processed by amateur astronomers uncovers more features\footnote{{Even though not all amateur astronomers recover the two distinct features of the double loop.}}. As a benchmark test the faintest stars in the different images could be compared. In our images, we reach faint stars up to $g=26.0$\,mag.} 

Furthermore, there is a puzzle concerning the exposure time. For example, the re-observation of NGC\,5907 by the astro-photographer Stefen Binnewies, using a 0.6 meter telescope and $24\times600$\,s exposures, i.e. 4 hours in total in the $L$-band, clearly shows part of the double loop.  {It is highly unlikely that our  $86\times300$\,s exposures, i.e. 7.2 hours in total in the same band, while having a larger collecting area of more than a factor of 5, is unable to detect the double loop\footnote{\om{However, a comparison between the exposure times of different telescopes is strictly speaking only valid if the CCD sensitivity, the pixel size, and the observation site are taken into account.}}.}
Even more surprising is that the brighter parts of the missing double loop appear as bright in the astro-photographer's images\footnote{see their stacked and enhanced image on http://www.capella-observatory.com/ImageHTMLs/Galaxies/NGC5907Prim.htm} as the here detected parts of the stream. If the apparent surface brightnesses are similar, then we should be able to detect both features too, which we apparently do not. This again points towards a problem {that the brightness in their post-processing is not preserved.}

\om{Having discussed all this, is there still a way to reconcile all observations? A difference between professional and amateur systems could be the spectral sensitivities of the CCDs and the transmission curves of the filters. In this case, different features would appear more or less prominent{, e.g. if the emission is coming from [OII] at 372.7\,nm, which is just outside of the typical SDSS $g$-band transmission curve}. However, this would mean that by digging deeper or changing the filters{ -- e.g. by employing SDSS $u$-band --} we should eventually be able to see the double loop if it is there.
}

To summarize this letter, we have acquired deep, high quality data with a 1.4 meter 
telescope, reaching a depth of 29.7\,mag arcsec$^{-2}$ in the $g$-band, and find no signs of the iconic double loop around NGC\,5907. Rather, we find a single knee-shaped stream as independently found by the Dragonfly team. \om{These results emphasize the need for an independent confirmation of the reliability of the detected low-surface brightness features by the usage of different telescopes, carefully designed observing strategies, and consistent scientific processing techniques.}

This work used following scientific software: Astrometry.net \citep{2010AJ....139.1782L},
the DAOPHOT2 suite \citep{1987PASP...99..191S}, and IRAF \citep{1993ASPC...52..173T}; as well as following python3 packages: Astropy \citep{astropy:2013}, ccdproc \citep{2015ascl.soft10007C}, and sep \citep{2016JOSS....1...58B}, which is a python implementation of  Source Extractor \citep{1996A&AS..117..393B}.

\begin{acknowledgements}
{We thank the referee for the constructive report, which helped to clarify and improve the manuscript.}
O.M. is grateful to the Swiss National Science Foundation for financial support. The authors thank Pieter van Dokkum for fruitful discussions about the observations and providing the Dragonfly data shown in Figure\,\ref{fig:comparison}. \om{The authors also thank Marina Rejkuba and Rodrigo Ibata for valuable comments.} This work was supported by the Ministry of Education, Science and Technological Development of the Republic of Serbia (MESTDRS) through project no.~176021, ``Visible and Invisible Matter in Nearby Galaxies: Theory and Observations''. We acknowledge the financial  support by the European Commission through project BELISSIMA (BELgrade  Initiative  for Space  Science,  Instrumentation  and  Modelling  in Astrophysics,  call  FP7-REGPOT-2010-5,  contract  No.  256772), which was used to procure the Milankovi\'c 1.4 meter telescope with the support from the MESTDRS.
We thank the technical operators at the Astronomical Station  Vidojevica (ASV), Miodrag Sekuli\'c and Petar Kosti\'c for their excellent work.
\end{acknowledgements}

\bibliographystyle{aa}
\bibliography{aanda}

   \end{document}